\documentclass[10pt,twocolumn,letterpaper]{article}

\usepackage{iccv}
\usepackage{times}
\usepackage{epsfig}
\usepackage{graphicx}
\usepackage{amsmath}
\usepackage{amssymb}

\usepackage{url}            
\usepackage{booktabs}       
\usepackage{amsfonts}       
\usepackage{nicefrac}       
\usepackage{microtype}      

\usepackage{microtype}
\usepackage{graphicx}
\usepackage{subfigure}
\usepackage{booktabs} 
\usepackage{times,epsfig}
\usepackage{bm}
\usepackage{amsfonts,amssymb,amsmath,amsth
m}
\usepackage{threeparttable}
\usepackage{array}
 \usepackage{multirow}
\usepackage{threeparttable}
\usepackage{extarrows}
\usepackage{amsmath}
\usepackage[table,xcdraw]{xcolor}
\usepackage{makecell}

\usepackage{algorithm,  algpseudocode}
\let\oldReturn\Return
\renewcommand{\Return}{\State\oldReturn}

\usepackage{xcolor}

\def\red#1{\textcolor{red}{#1}}
\def\blue#1{\textcolor{blue}{#1}}
\def\orange#1{\textcolor{orange}{#1}}

\newtheorem{theorem}{Theorem}

\newtheorem{assumption}{Assumption}

\usepackage{amsmath}

\newcommand{\x}{\bm{x}}

\newcommand{\z}{\bm{z}}

\newcommand{\C}{\mathcal{C}}
\newcommand{\F}{\mathcal{F}}
\newcommand{\X}{\mathcal{X}}
\newcommand{\Y}{\mathcal{Y}}

\newcommand{\tabincell}[2]{\begin{tabular}{@{}#1@{}}#2\end{tabular}} 

\long\def\comment#1{}

\usepackage{multirow}

\usepackage[pagebackref=true,breaklinks=true,letterpaper=true,colorlinks,bookmarks=false]{hyperref}

\iccvfinalcopy 

\def\iccvPaperID{2786} 

\ificcvfinal\fi

\begin{document}

\title{Boosting Black-Box Attack with Partially Transferred Conditional Adversarial Distribution}


\author{%
	Yan Feng$^{1}$, 
Baoyuan Wu$^{2,3}$, Yanbo Fan$^{4}$,
Li Liu$^{2,3}$,
Zhifeng Li$^4$, Shutao Xia$^{1}$\\
$^1$Tsinghua Shenzhen International Graduate School, Tsinghua University, China\\
$^2$School of Data Science, The Chinese University of Hong Kong, Shenzhen, China\\
$^3$Shenzhen Research Institute of Big Data, Shenzhen, China\\
$^4$Tencent AI Lab, China\\
\small{\texttt{wubaoyuan1987@gmail.com}; \texttt{xiast@sz.tsinghua.edu.cn}}\\
}

\maketitle
\ificcvfinal\thispagestyle{empty}\fi

\begin{abstract}

This work studies black-box adversarial attacks against deep neural networks (DNNs), where the attacker can only access the query feedback returned by the attacked DNN model, while other information such as model parameters or the training datasets are unknown. 
One promising approach to improve attack performance is utilizing the adversarial transferability between some white-box surrogate models and the target model (i.e., the attacked model). 
However, due to the possible differences on model architectures and training datasets between surrogate and target models, dubbed ``surrogate biases", the contribution of adversarial transferability to improving the attack performance may be weakened.
To tackle this issue, we innovatively propose a black-box attack method by developing a novel mechanism of adversarial transferability, which is robust to the surrogate biases. 
The general idea is transferring partial parameters of the conditional adversarial distribution (CAD) of surrogate models, while learning the untransferred parameters based on queries to the target model, to keep the flexibility to adjust the CAD of the target model on any new benign sample. \comment{Consequently, the possible negative effect due to surrogate biases of model architectures and training samples could be mitigated. Besides, since the class labels are not involved in CAD, it is also robust to the surrogate bias of class labels.} Extensive experiments on benchmark datasets and attacking against real-world API demonstrate the superior attack performance of the proposed method.

\end{abstract}

\vspace{-1em}
\section{Introduction}
\vspace{-.5em}


It has been well known \cite{BiggioCMNSLGR13, GoodfellowSS14} that adversarial examples are serious threats to deep neural networks (DNNs). 
Existing adversarial attacks can be generally partitioned into two main categories.
The first category is {\it white-box attack} \cite{GoodfellowSS14}, where the attacker can access parameters of the attacked DNN model. 
The second one is {\it black-box attack} \cite{DongSWLL0019}, where the attacker can only access the query feedback returned by the attacked model, while model parameters are unknown to the attacker.
Since it is difficult to access model parameters in real-world scenarios, black-box attack is more practical, and it is also the main focus of this work.

If only utilizing the query feedback, it is difficult to achieve high attack success rate under limited query budgets. One promising approach to improve the attack performance, including attack success rate and query efficiency, is utilizing the adversarial transferability \cite{DongLPS0HL18, DongPSZ19, Wu0X0M20} between some white-box surrogate models and the target model (\ie, the attacked model). 
Many adversarial transferabilities have been proposed in existing works, such as the gradient \cite{ChengDPSZ19, GuoYZ19}, or the projection from a low-dimensional space to the original sample space \cite{Huang020}, \etc. 
These transferabilities have shown positive contributions to improving the attack performance in some black-box attack scenarios, especially in the \textbf{closed-set scenario}, where the training dataset of the target model is known to the attacker. 
However, their effects may be significantly influenced by the differences between surrogate and target models.
More precisely, architectures between surrogate and target models may be different, probably leading to different feedback to the same query. 
Secondly, under the practical scenario of \textbf{open-set black-box attack} where the training dataset is unknown to the attacker, even using the same architecture, different training sets (including samples and class labels) will also lead to different parameters. 
We generally summarize the differences, caused by architectures and training datasets between surrogate and target models as \textbf{surrogate biases}. 
If the biases are too large, the transferred information may mislead the search of adversarial perturbation for attacking the target model, causing the degradation of the contribution of adversarial transferability to improving the attack performance  (as demonstrated later in Sec. \ref{sec:experiments:openset}). 



To mitigate the above issue, the transferred term should be {not only informative but also} robust to surrogate biases. To this end, we focus on the \textbf{conditional adversarial distribution (CAD)} (\ie, the distribution of adversarial perturbations conditioned on benign examples). 
If the transferred CAD accurately fits the target model, it will be helpful to search successful adversarial perturbations for attacking the target model. 
Besides, note that CAD is independent with class labels, thus transferring CAD will be robust to the surrogate bias of training class labels.
However, CAD can be influenced by the biases of model architectures and training samples. Thus, we propose a novel transfer mechanism that only partial parameters of CAD are transferred, while the remaining parameters are learned according to the query feedback returned by the target model on the attacked benign sample.
Consequently, the CAD of the target model conditioned on any new benign sample could be flexibly adjusted, such that the possible negative effect due to surrogate biases of architectures and training samples could be mitigated. 
One remaining important issue is how to accurately model the CAD. Here we adopt the conditional generative flow model, called {\bf c-Glow} \cite{abs-1905-13288}, whose general idea is invertibly mapping a simple distribution (\eg, Gaussian distribution) to a complex distribution through an invertible network, as shown in Fig.~\ref{fig: pipeline}(a). 
c-Glow has shown powerful ability of capturing complex data distributions \cite{abs-1905-13288}, and we believe that it is capable enough to capture the CAD. 
{To the best of our knowledge, this is the first work to use c-Glow to approximate the CAD.}
Besides, we develop an efficient training algorithm of the c-Glow model based on randomly sampled perturbations, rather than costly generated adversarial perturbations, such that the CAD of surrogate models can be efficiently and accurately approximated. 
Extensive experiments are conducted to verify the effectiveness of the proposed attack method, including black-box attack scenarios of both  \textbf{closed-set} and \textbf{open-set}  on benchmark datasets, as well as the attack against real-world API.

In summary, the main contributions of this work are threefold.
{\bf 1)} We propose an effective and efficient black-box attack method by designing a novel adversarial transfer mechanism that only partial parameters of the conditional adversarial distribution are transferred, which is robust to surrogate biases between surrogate and target models. 
{\bf 2)} We are the first to approximate the CAD by the c-Glow model, and design an efficient training algorithm based on randomly sampled perturbations.
{\bf 3)} Extensive experiments demonstrate the superiority of the proposed attack method to several state-of-the-art (SOTA) black-box methods by improving attack success rate and query efficiency simultaneously.

\vspace{-.5em}
\section{Related Work}
\vspace{-.5em}

Adversarial attack has been well studied in recent years. Please refer to \cite{AkhtarM18} for a detailed survey. In this section, we mainly discuss the related works of black-box adversarial attack methods, including decision-based and score-based adversarial attacks.

\noindent {\bf Decision-based Adversarial Attacks.} 
For decision-based attacks, an attacker can only acquire the output label of the target model. Boundary Attack \cite{BrendelRB18} first studies the problem by randomly sampling candidate perturbations following the normal distribution, and the perturbation with the lower objective is updated as the new solution. 
An evolution based search method \cite{DongSWLL0019} utilized the history queries to approximate a Gaussian distribution as the search distribution. 
\cite{ChengLCZYH19} formulated the decision-based attack problem as a continuous optimization by alternatively optimizing the perturbation magnitude and perturbation direction.
This method was further accelerated in \cite{ChengSCC0H20} by only estimating the sign of gradient.
HopSkipJumpAttack \cite{abs-1904-02144} developed an iterative search algorithm by utilizing binary information at the decision boundary to estimate the gradient. 
It is further improved in \cite{abs-2005-14137} by learning a more representative subspace for perturbation sampling.
Based on the observation of the low curvature of the decision boundary around adversarial examples, a gradient approximation method was proposed in \cite{LiuMF19} based on the gradients of neighbour points. GeoDA
\cite{abs-2003-06468} locally approximated the decision boundary with a hyper-plane, and searched the closest point on the hyper-plane to the benign input as the perturbation.

\vspace{-.4em}
\noindent {\bf Score-based Adversarial Attacks.}
There are generally three sub-categories of score-based black-box attacks, including  {\it transfer-based}, {\it query-based} and {\it query-and-transfer-based} attacks. 
\textbf{1)} {\bf Transfer-based methods} attempt to generate adversarial perturbations utilizing the information of surrogate white-box models. For example, it was proposed in \cite{PapernotMG16} to firstly train a surrogate white-box model with a dataset labeled by querying the target model, then utilize the gradient of the trained surrogate model to generate adversarial perturbations to attack the target model. 
Adversarial perturbations was found  in \cite{LiuCLS17}  to achieve better attack performance when generated on an ensemble of source models. Recently, \cite{inkawhich2020perturbing} proposed to perturb across the intermediate feature space, rather than focus solely on the output layer of the source models, so as to improve the transferability of the generated adversarial examples. 
Although transfer-based attack methods are very efficient, the attack success rate is often lower than query-based attack methods.
\textbf{2)} {\bf Query-based methods} solve the black-box optimization by iteratively querying the target model. 
SimBA \cite{GuoGYWW19} randomly sampled a perturbation from a predefined orthonormal basis, and then either added or subtracted this perturbation to the attacked image.
Natural evolution strategy (NES) \cite{nes-jmlr-2014,WierstraSPS08}  was adopted in \cite{IlyasEAL18} to minimize a continuous expectation of the black-box objective function based on a search distribution.
Bandit \cite{IlyasEM19} improved the NES method by incorporating data and temporal priors into the gradient estimation. 
SignHunter \cite{Al-DujailiO20} adopted the gradient sign rather than the gradient as the search direction.
Query-based methods often achieve better attack performance than transfer-based methods, but require more queries. 
\textbf{3)} {\bf Query-and-transfer-based methods} try to take advantage of both transfer-based and query-based methods, to achieve high attack success rate and high query efficiency simultaneously.  
%
The general idea is firstly learning some types of priors from surrogate models, then incorporating these priors into the query-based method to guide the attack procedure for the target model. 
For example, the prior used in $\mathcal{N}$ATTACK \cite{LiLWZG19} is the mean parameter of the Gaussian search distribution in NES, which is learned using a regression neural network trained based on surrogate models. 
AdvFlow \cite{advflow} assumes that the marginal distributions of benign examples and adversarial examples are similar, to generate inconspicuous adversaries.  
The prior used in Square attack \cite{ACFH2020square} is that it is more likely to find an adversarial perturbation at the boundary of the feasible set of allowed perturbations. 
Methods in \cite{ChengDPSZ19} and \cite{GuoYZ19} utilized the gradient of surrogate models as the gradient prior.
TREMBA \cite{Huang020} treated the projection from a low-dimensional space to the original space as the prior, such that the perturbation could be search in the low-dimensional space.
The hybrid method \cite{abs-1908-07000} proposed to initialize the attack with adversarial perturbations from the surrogate models and update surrogate models using the feedback of the target model. LeBA \cite{yang2020learning} also proposed to update the surrogate models to approximate the target models\comment{Instead of directly re-train the surrogate models based on the attack data and prediction of target models as \cite{abs-1908-07000}}, where a high order computation graph was built for updating the surrogate models in both forward and backward pass.
Besides, recently a few algorithms were specially developed to handle the open-set black-box attack scenario (also called data-free black-box attack in \cite{zhou2020dast,huan2020data}). However, the DaST method \cite{zhou2020dast} required massive queries to train a surrogate model, which doesn't satisfy the goal of improving attack performance under limited query budgets. 
The DFP method \cite{huan2020data} assumed that the target model is fine-tuned based on a white-box pre-trained model, and the attack success rate is very low. 
In contrast, our method also belongs to the last category and is applicable to both closed-set or open-set scenarios, but shows superior attack performance in both scenarios, due to the partial transfer mechanism.

\vspace{-.6em}
\section{Method}
\label{sec: method}


\subsection{Problem Formulation of Black-Box Attack}
\label{sec: subsec adversarial attack}

We denote a classification model $\mathcal{F}: \mathcal{X} \rightarrow \mathcal{Y}$, with $\mathcal{X}$ being the input space, $n=|\X|$ indicating the dimension of the input space, and $\mathcal{Y}$ being the output space. 
Given a benign example $\x \in \X$ and its ground-truth label $y \in \Y$,
$\mathcal{F}(\x, y) \in [0,1]$ indicates the classification score $w.r.t.$ the $y$-th label. 
In this work, we adopt the logit as the classification score. 
The goal of {\it adversarial attack} is finding a small perturbation $\bm{\eta}$ within a $\ell_p$-ball, \ie, $\mathbb{B}_{\epsilon} = \{ \bm{\eta} | \bm{\eta} \in \mathbb{R}^{n}, ~ \|\bm{\eta}\|_{p} \leq \epsilon \}$ ($\epsilon > 0$ being an attacker defined scalar, which will be specified in Sec. \ref{sec: setting}), 
such that the prediction of $\x + \bm{\eta}$ is different from the prediction of $\x$.
Specifically, the attack problem can be formulated as minimizing $\mathcal{L}_{adv}$
\comment{
\vskip -0.2in
\begin{flalign}
   \min\limits_{\bm{\eta}} & ~~\mathcal{L}_{adv}(\bm{\eta}, \bm{x}, y) = \mathbb{I}\big(\bm{\eta} \in \mathbb{B}_{\epsilon} \big) + 
   \max\big(0, \triangle \big),
    \label{eq: untargeted attack}
\end{flalign}
\vskip -0.1in
}
\begin{flalign}
\vspace{-0.4em}
   \mathcal{L}_{adv}(\bm{\eta}, \bm{x}, y) = \mathbb{I}\big(\bm{\eta} \in \mathbb{B}_{\epsilon} \big) + 
   \max\big(0, \triangle \big),
    \label{eq: untargeted attack}
\end{flalign}
where $\triangle = \F(\bm{x} + \bm{\eta}, y) - \max \limits_{j \neq y} \F(\bm{x} + \bm{\eta}, j)$ for the {\it untargeted attack}, while 
$\triangle = \max\limits_{j \neq t} \F(\bm{x} + \bm{\eta}, j) - \F(\bm{x} + \bm{\eta}, t)$ for the {\it targeted attack} with $t \in \Y$ being the target label. 
$\mathbb{I}(a) = 0$ if $a$ is true, otherwise $\mathbb{I}(a)$ = $+\infty$, which enforces that the perturbation $\bm{\eta}$ is within the range $\mathbb{B}_{\epsilon}$.
Note that $\mathcal{L}_{adv}$ is {\it non-negative}, and if $0$ is achieved, then the corresponding $\bm{\eta}$ is a successful adversarial perturbation. 

Here we consider a practical and challenging scenario that parameters of $\F$ are unknown to the attacker, while only the classification score $\mathcal{F}(\x, y)$ is returned by querying $\F$, dubbed \textbf{score-based black-box attacks}. Furthermore, if the training dataset of $\F$ is known to the attacker, then it is called \textit{closed-set attack scenario}, otherwise called \textit{open-set attack scenario}. 
The goal of black-box attacker is to find a successful adversarial perturbation $\bm{\eta}$ (\ie, $\mathcal{L}_{adv}(\bm{\eta}, \bm{x}, y)=0$) under limited query budgets. In other words, a good attack algorithm should achieve \textit{high attack success rate} (ASR) and \textit{high attack efficiency} (\ie, fewer queries) simultaneously.  
To this end, one promising approach is utilizing both the query feedback returned by the target model and adversarial transferability from some white-box surrogate models, dubbed \textit{query-and-transfer-based} attack method. 
The effect of transferability is related to the differences between surrogate and target models, including model architectures, training samples, as well as training class labels, as these information of the target model is unknown to the attacker in practical scenarios, especially in the open-set scenario. These differences are generally called \textbf{surrogate biases}, which may cause negative transfer to harm the attack performance. 

To mitigate the possible negative effect from adversarial transferbility, here we propose a novel transfer mechanism that is robust to surrogate biases. The general idea is partially transferring the CAD of surrogate models, while keeping the flexibility to adjust the CAD according to queries to the target model.  
In the following, we will firstly introduce the modeling of CAD using the c-Glow model in Sec. \ref{sec: subsec c-glow model}; then, we will present the attack method utilizing the proposed transfer mechanism, called \textbf{$\bm{\mathcal{CG}}$-ATTACK}, in Sec. \ref{method}.

\vspace{-.2em}
\subsection{Modeling Conditional Adversarial Distribution}
\label{sec: subsec c-glow model}
\vspace{-.3em}

\subsubsection{Conditional Glow Model}
\begin{figure}[t] 
\vspace{-.5em}
\begin{center}
\centerline{\includegraphics[width=1\linewidth]{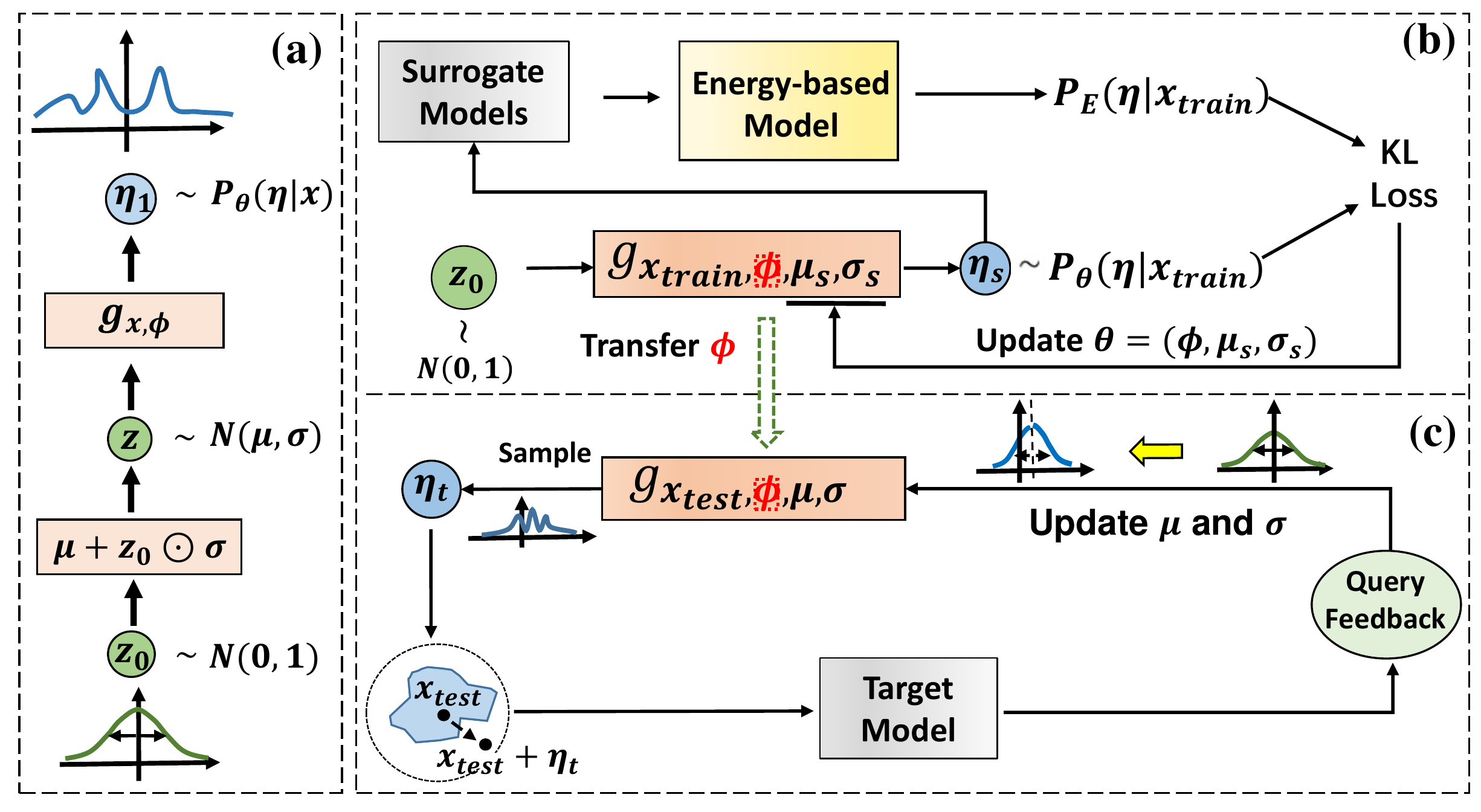}}
\vspace{-.3em}
\caption{Overall pipeline of our method. {\bf (a)} The general structure of the c-Glow model, which maps the simple normal distribution to the \textit{CAD}. {\bf (b)} The efficient training method of c-Glow on surrogate white-box DNN models. {\bf (c)} The proposed black-box attack method $\mathcal{CG}$-ATTACK, which transfers the mapping parameter $\bm{\phi}$ of the c-Glow model trained on surrogate DNN models. }
\label{fig: pipeline}
\vspace{-3em}
\end{center}
\end{figure}

The c-Glow model was recently proposed in \cite{abs-1905-13288} to learn the complex posterior probability in structured output learning. 
It can generate an invertible mapping between one random variable $\bm{\eta}$ and another random variable $\z$ that follows a simple distribution (\eg, Gaussian distribution), given the condition $\x$.
c-Glow can be formulated as an inverse function $g_{\x, \bm{\phi}}: \z \rightarrow \bm{\eta}$, and there exists 
$g_{\x, \bm{\phi}}^{-1}: \bm{\eta} \rightarrow \z$, with $\bm{\phi}$ indicating the mapping parameter. 
In the scenario of adversarial attack, the condition variable $\x \in \X$ is the benign example, and $\bm{\eta} \in \mathbb{R}^{|\X|}$ represents the perturbation variable.
$g_{\bm{\phi}, \x}$ can be further decomposed to the composition of $M$ inverse functions \cite{abs-1905-13288}, as follows:
\begin{flalign}
\vspace{-.7em}
\bm{\eta}  = g_{\x, \bm{\phi}}(\z) = g_{\bm{x}, \bm{\phi}_1}( g_{\bm{x}, \bm{\phi}_{2}}( ... (g_{\bm{x}, \bm{\phi}_M}(\z) ) ...)), 
\label{eq: sequential product from z to eta}
\vspace{-.7em}
\end{flalign} 
where $\bm{\phi}$ is specified as $(\bm{\phi}_1, \ldots, \bm{\phi}_M)$, and 
$\bm{\phi}_i$ indicates the parameter of $g_{\bm{x}, \bm{\phi}_i}(\cdot)$. 
The c-Glow model can be represented by a neural network with $M$ layers ($M$ is set to 3). 
Each layer consists of a conditional actnorm module, followed by a conditional $1 \times 1$ convolutional module and a conditional coupling module. 
A general structure of c-Glow is shown in Fig. \ref{fig: pipeline}(a). The detailed \comment{definition of $g_{\bm{x}, \bm{\phi}_i}(\cdot)$ and the corresponding structure} description of the c-Glow model will be presented in Sec. 1 of the {\bf Supplementary Material}.

\vspace{-1.0em}
\subsubsection{Approximating \textit{CAD} by c-Glow}
\vspace{-0.7em}

Instead of modeling the marginal distribution  $\mathcal{P}_{\bm{\theta}}(\bm{\eta})$, here we propose to utilize the powerful capability of c-Glow to approximate the \textit{CAD} (\ie, $\mathcal{P}_{\bm{\theta}}(\bm{\eta}|\bm{x})$) in the task of adversarial attack.
%
Based on the mapping from the latent variable $\z$ to the perturbation variable $\bm{\eta}$ (\ie, Eq. (\ref{eq: sequential product from z to eta})), we derive a mathematical formulation of  $\mathcal{P}_{\bm{\theta}}(\bm{\eta}|\bm{x})$. 
Specifically, we first set $\z = \bm{\mu} + \bm{\sigma} \odot \z_0$ with $\z_0 \sim \mathcal{N}(\bm{0}, \mathbf{I})$, where $\odot$ is the entry-wise product and $\mathbf{I}$ indicates the identity matrix. 
Then, utilizing {\it the change of variables} \cite{Tabak2010Density} of Eq. (\ref{eq: sequential product from z to eta}), the conditional likelihood of $\bm{\eta}$ given $\x$ can be formulated as
\begin{small}
\begin{equation}  
    \log \mathcal{P}_{\bm{\theta}}(\bm{\eta}|\bm{x}) = \log \mathcal{P}_{\bm{0, 1}}(\bm{z}_0) + \sum\limits_{i=1}^{M+1} \log \bigg|\det \bigg(\frac{\partial g^{-1}_{\x, \bm{\phi}_{i}}(\bm{r}_{i-1})}{\partial \bm{r}_{i-1}} \bigg) \bigg|,
\label{eq: prob of c-glow}
\end{equation}
\end{small}
where $\bm{\theta} = (\bm{\phi}, \bm{\mu}, \bm{\sigma})$,
$\bm{r}_i = g^{-1}_{\bm{\phi}_i, \x}(\bm{r}_{i-1})$, $\bm{r}_0 = \bm{\eta}$, $\bm{r}_M = \bm{z}$ and $\bm{r}_{M+1} = \bm{z}_0$, with $i$ indicating the index of the $i$-th inverse function in c-Glow. 
$\det(\cdot)$ indicates the determinant of a matrix, and
$\mathcal{P}_{\bm{0, 1}}(\cdot)$ indicates the probability density function of the multi-variant normal distribution $\mathcal{N}(\bm{0}, \mathbf{I})$. 
For simplicity, in Eq. (\ref{eq: prob of c-glow}) we treat the transformation $\z = \bm{\mu} + \bm{\sigma} \odot \z_0$ as the $M+1$ layer of the c-Glow model, \ie, $g_{\x, \bm{\phi}_{M+1}}(\z_0) = \bm{\mu} + \bm{\sigma} \odot \z_0$ with $\bm{\phi}_{M+1} = (\bm{\mu}, \bm{\sigma})$, which is also invertible, but is independent on $\x$. Thus, we have 
 $\bm{\eta}= g_{\x, \bm{\theta}}(\z_0) = g_{\x, \bm{\phi}}(\z)$. 

\vspace{-.7em}
\subsubsection{Learning of the c-Glow Model}
\label{sec: subsec learning c-glow}
\vspace{-.3em}

In \cite{abs-1905-13288}, the parameter $\bm{\theta}$ of c-Glow is learned via maximum likelihood estimation (\ie, $\max_{\bm{\theta}} \log \mathcal{P}_{\bm{\theta}}(\bm{\eta}|\bm{x})$). 
However, it may not be a suitable choice for approximating the \textit{CAD}, because it generally requires massive
adversarial perturbations, when there are multiple layers in
the adopted c-Glow model. Meanwhile, the generation of these adversarial perturbations is very time consuming. 
Recall that our work is to transfer the mapping parameters learned by the c-Glow in white-box attack scenario to the black-box attack scenario. To tackle the above challenge, we first present a novel learning method based on surrogate white-box models.

\vspace{.1em}
\noindent {\bf Energy-based Model}. By utilizing the adversarial loss $\mathcal{L}_{adv}(\bm{\eta}, \x)$, we define an energy-based model \cite{HintonOWT06} to capture the distribution of $\bm{\eta}$ around $\x$, as follows
\begin{small}
\vspace{-.5em}
\begin{flalign} 
\mathcal{P}_{E}(\bm{\eta}| \x) = \frac{\exp\big(-\lambda \cdot \mathcal{L}_{adv}(\bm{\eta}, \x) \big)}{\int_{\bm{\eta} \in \mathbb{B}_{\epsilon} } \exp\big(-\lambda \cdot \mathcal{L}_{adv}(\bm{\eta}, \x) \big) d \bm{\eta} }.
\end{flalign}
\vspace{-.5em}
\end{small}

Note that given the classification model $\F$, the normalization term (\ie, the denominator) is an intractable constant. Thus, we simply omit it hereafter, and set 
\begin{small}
\vspace{-.5em}
\begin{flalign}
\log \mathcal{P}_{E}(\bm{\eta}| \x) \approx -\lambda \cdot \mathcal{L}_{adv}(\bm{\eta}, \x),
\label{assumption}
\end{flalign}
\end{small}
where $\lambda$ is a positive hyper-parameter, which will be specified in {Sec. 5 of the \textbf{Supplementary Material}}.
%
Note that the distributions of both untargeted and targeted adversarial perturbations can be formulated by Eq. (\ref{assumption}), by specifying $\triangle$ to the corresponding format in $\mathcal{L}_{adv}(\bm{\eta}, \x)$ (see Eq. (\ref{eq: untargeted attack})).

In practice, we randomly sample a large number of perturbations within the neighborhood $\mathbb{B}_{\epsilon}$ around each benign example $\x$, then feed the perturbed example $\x+\bm{\eta}$ into the attacked model to obtain the values of $\log \mathcal{P}_{E}(\bm{\eta}| \x)$. 
Note that we only need to sample perturbations within $\mathbb{B}_{\epsilon}$, as the values of $\log \mathcal{P}_{E}(\bm{\eta}| \x)$ for outer perturbations are negative infinity (see Eq. (\ref{eq: untargeted attack})), which are useless for learning.

\vspace{0.2em}
\noindent {\bf Minimization of KL divergence.}
Given $\mathcal{P}_{E}(\bm{\eta}| \x)$ defined in Eq.~(\ref{assumption}), we propose to learn the parameter $\bm{\theta}$ of the c-Glow model by minimizing the KL divergence \cite{kullback1951} between $\mathcal{P}_{E}(\bm{\eta}| \x)$ and $\mathcal{P}_{\bm{\theta}}(\bm{\eta}| \x)$. 
The rationale behind is that if the adversarial probabilities for any perturbation evaluated by both $\mathcal{P}_{E}(\bm{\eta}| \x)$ and $\mathcal{P}_{\bm{\theta}}(\bm{\eta}| \x)$ are similar, then the learned $\mathcal{P}_{\bm{\theta}}(\bm{\eta}| \x)$ can be considered as an good approximation to the real adversarial distribution.
Without loss of generality, we consider one benign example $\x$, then the learning of $\bm{\theta}$ is formulated as the minimization of the following objective,
\begin{small}
\vspace{-0.2em}
\begin{align} \label{equ: kl divergence}
     \mathcal{L} &=  
    \mathbb{E}_{\mathcal{P}_E(\bm{\eta}|\x)}
    \left[\log \frac{\mathcal{P}_E(\bm{\eta}|\x)}{\mathcal{P}_{\bm{\theta}}(\bm{\eta}|\x)} \right].
\end{align}
\vspace{-0.2em}
\end{small}
\vskip -0.1in
We adopt the gradient-based method to optimize this problem, and the gradient of $\mathcal{L}$ $w.r.t.$ $\bm{\theta}$ is presented in Theorem \ref{t1}. 
Due to the space limit, the proof of Theorem \ref{t1} will be presented in the Sec. 2 of the {\bf Supplementary Material}.
Note that each term within the expectation in Eq. (\ref{therem}) is tractable, thus $\nabla_{\bm{\theta}} \mathcal{L}$ can be easily computed. In practice, $K$ instantiations of $\z_0$ are sampled from $\mathcal{N}(\bm{0}, \mathbf{I})$, then $\nabla_{\bm{\theta}} \mathcal{L}$ is empirically estimated as the average value over these $K$ instantiations. 
The general structure of the proposed learning method is presented in Fig. \ref{fig: pipeline}(b). 
\vspace{-.5em}
\begin{theorem} \label{t1}
Utilizing $\bm{\eta}= g_{\x, \bm{\theta}}(\z_0)$ and $\z_0 \sim \mathcal{N}(\bm{0}, \mathbf{I})$ defined in Sec. \ref{sec: subsec c-glow model}, as well as Eq. (\ref{assumption}), and defining the term $D(\bm{\eta}, \bm{x}) = \log \frac{\mathcal{P}_E(\bm{\eta}| \bm{x})}{\mathcal{P}_{\bm{\theta}}(\bm{\eta}| \bm{x})}$, then the gradient of $\mathcal{L}$ w.r.t. $\bm{\theta}$ can be computed as follows
\begin{small}
\begin{align} \label{therem}
\vspace{-.3em}
    \nabla_{\bm{\theta}} \mathcal{L}  &= 
    - \mathbb{E}_{\z_0 \sim \mathcal{N}(\bm{0}, \mathbf{I})}  
    \big[
    \frac{\exp^{-\lambda \cdot \mathcal{L}_{adv}(\bm{\eta}, \bm{x})}}{\mathcal{P}_{\bm{\theta}}(\bm{\eta} | \bm{x})}
    \cdot
    \nabla_{\bm{\theta}}g_{\bm{x},\bm{\theta}}(\bm{z_0})
     \nonumber
    \\
    &~~~~~
    \cdot 
    \nabla_{\bm{\eta}} D(\bm{\eta}, \bm{x})^\top \big|_{\bm{\eta} = g_{\bm{x},\bm{\theta}}(\bm{z}_0)}
    \big],
\end{align}{}
\end{small}
\comment{
\begin{align} \label{therem}
\vspace{-.4em}
    \nabla_{\bm{\theta}} \mathcal{L}  &= 
    - \mathbb{E}_{\z_0 \sim \mathcal{N}(\bm{0}, \mathbf{I})}  
    \big[
    \exp^{D(\bm{\eta}, \bm{x})} 
     \cdot
    \nabla_{\bm{\theta}}g_{\bm{x},\bm{\theta}}(\bm{z_0})
    \\
    &~~~~~
    \cdot 
    \nabla_{\bm{\eta}} D(\bm{\eta}, \bm{x})^\top \big|_{\bm{\eta} = g_{\bm{x},\bm{\theta}}(\bm{z}_0)}
    \big],
     \nonumber
    \\
    &= 
    - \mathbb{E}_{\z_0 \sim \mathcal{N}(\bm{0}, \mathbf{I})}  
    \big[
    \frac{\exp^{-\lambda \cdot \mathcal{L}_{adv}(\bm{\eta}, \bm{x})}}{\mathcal{P}_{\bm{\theta}}(\bm{\eta} | \bm{x})}
    \cdot
    \nabla_{\bm{\theta}}g_{\bm{x},\bm{\theta}}(\bm{z_0})
     \nonumber
    \\
    &~~~~~
    \cdot 
    \nabla_{\bm{\eta}} D(\bm{\eta}, \bm{x})^\top \big|_{\bm{\eta} = g_{\bm{x},\bm{\theta}}(\bm{z}_0)}
    \big],
    \nonumber
\end{align}{}
}
where $\nabla_{\bm{\eta}} D(\bm{\eta}, \bm{x}) = \nabla_{\bm{\eta}} [ -\lambda  \mathcal{L}_{adv}(\bm{\eta}, \bm{x}) - \log \mathcal{P} _{\bm{\theta}}(\bm{\eta} | \bm{x})]$.
\end{theorem}

\vspace{-.5em}
\begin{algorithm}[b]
\vspace{-.1em}
\footnotesize
\caption{The proposed $\mathcal{CG}$-ATTACK method with CMA-ES being the basic algorithm.}
\label{alg:evolution}
\begin{algorithmic}[1]
\Require The black-box attack objective $\mathcal{L}_{adv}(\cdot, \x)$ with the benign input $\x$, the ground-truth label $y$ or the target label $t$, population size $k$, surrogate white-box models, training set $\mathcal{D}$ of the surrogate models, the maximal number of queries $T$, the downsampling ratio $r$.
\State Pretrain the c-Glow model in the $r$-DCT subspace of $\mathcal{D}$ based on surrogate models, and obtain the parameters $\bm{\phi}, \bm{\mu}_s, \bm{\sigma}_s$;
\State Initialize $\bm{\mu} = \bm{\mu}_s, \bm{\sigma} = \mathbf{I}$, and  initialize other parameters in the standard CMA-ES algorithm;
\For {$t = 1$ to $T$}
\State Sample $k$ perturbations $\bm{\eta}_1, ..., \bm{\eta}_k \sim \mathcal{P}_{(\bm{\phi}, \bm{\mu}, \bm{\sigma})}(\bm{\eta} | \x)$;
\State Upsample the perturbations $\bm{\eta}_1, ..., \bm{\eta}_k$ with IDCT into the same size of $\x$, obtaining $\bm{\bar{\eta}}_1, ..., \bm{\bar{\eta}}_k$; 
\State Evaluate $\mathcal{L}_{adv}(\bm{\bar{\eta}}_1, \x), ..., \mathcal{L}_{adv}(\bm{\bar{\eta}}_k, \x)$;
\If {$\exists \bm{\bar{\eta}}_i, \mathcal{L}_{adv}(\bm{\bar{\eta}}_i, \x)=0$}
 \Return $\x + \bm{\bar{\eta}}_i$;
\EndIf
\State Update $\bm{\mu}, \bm{\sigma}$ and other parameters as did in the standard CMA-ES;
\EndFor
\end{algorithmic}
\vspace{-.2em}
\end{algorithm}

\vspace{-.2em}
\subsection{$\mathcal{CG}$-ATTACK} \label{method}

\noindent \textbf{Evolution-Strategy-based Attack Method.} 
Here we firstly give a brief introduction of evolution strategy (ES) \cite{HansenAA15, rechenberg1978evolutionsstrategien}, which has been widely used in black-box attacks, such as NES \cite{IlyasEAL18}, TREMBA \cite{Huang020}, $\mathcal{N}$ATTACK \cite{LiLWZG19}, \etc. 
The general idea of ES is introducing a search distribution to sample multiple perturbations $\bm{\eta}$, then these perturbations are fed into the target model to evaluate the corresponding objective values $\mathcal{L}_{adv}(\bm{\eta}, \x, y)$, which are then used to update the parameters of the search distribution based on some strategies (\eg, Natural ES \cite{nes-jmlr-2014,WierstraSPS08}, CMA-ES \cite{Hansen16a}). This process is repeated, until one successful adversarial perturbation is found (\ie, $\mathcal{L}_{adv}(\bm{\eta}, \x, y) = 0$). 
Instead of adopting the simple Gaussian distribution as the search distribution as did in TREMBA and $\mathcal{N}$ATTACK, here we specify the search distribution as the CAD modeled by the c-Glow model. 
As verified in experiments presented in {Sec. 7.4 of the \textbf{Supplementary Material}}, the c-Glow model can capture the CAD more accurately than the Gaussian model.  

\vspace{.3em}
\noindent \textbf{A Novel Transfer Mechanism of CAD.} 
One main challenge of the above ES-based black-box attack method is that there are significantly more parameters of the c-Glow model than Gaussian model, and it may require more queries to learn good parameters. Therefore, we resort to adversarial transferability, \ie, firstly learning the c-Glow model based on some white-box surrogate models using the learning algorithm in Sec. \ref{sec: subsec learning c-glow}, then transferring this learned c-Glow model to approximate the CAD of the target model. 
However, as mentioned in Sec. \ref{sec: subsec adversarial attack}, the CADs of surrogate and target models should be different, due to surrogate biases. The transfer of the whole c-Glow model may cause negative transfer to harm the attack performance. 
Thus, we propose a novel transfer mechanism that only transferring  mapping parameters $\bm{\phi}$ of the c-Glow model, while the remaining Gaussian parameters $\bm{\mu}$ and $\bm{\sigma}$ are learned based on queries to the target model, as shown in Fig. \ref{fig: pipeline}(c). 
The rationale behind this partial transfer is Assumption \ref{assumption 1}, which will be verified in {Sec. 7.2 of the \textbf{Supplementary Material}}. 
\vspace{-0.7em}
\begin{assumption}\label{assumption 1}
Given two c-Glow models learned for two DNN models, \ie, $g_{\x, \bm{\theta}_1}$ with $\bm{\theta}_1 = (\bm{\phi}_1, \bm{\mu}_1, \bm{\sigma}_1)$ and $g_{\x, \bm{\theta}_2}$ with $\bm{\theta}_2 = (\bm{\phi}_2, \bm{\mu}_2, \bm{\sigma}_2)$, we assume that their mapping parameters are similar, i.e., $\bm{\phi}_1 \approx  \bm{\phi}_2$.
\vspace{-0.7em}
\end{assumption}
We believe that this partial transfer mechanism has \textbf{two main advantages}. 
\textbf{1)} It keeps the flexibility to automatically adjust the CAD of the target model on the currently attacked sample $\x$, to mitigate the possible negative effect due to the surrogate biases from model architectures and training samples. 
\textbf{2)} Since this transfer is only related to the conditional probability $\mathcal{P}_{\bm{\theta}}(\bm{\eta} | \x)$, while independent with the marginal probability $\mathcal{P}(y)$, it is supposed to be robust to the surrogate bias of training class labels. 
Above advantages make the attack method utilizing this partial transfer mechanism more practical in real-world scenarios, especially in the open-set scenario. 
The attack method combining ES-based attack and this partial transfer mechanism based on the \textbf{C}onditional \textbf{G}low model is called $\bm{\mathcal{CG}}$-\textbf{ATTACK}, of which the general procedure is presented in Fig. \ref{fig: pipeline}. 

\vspace{.3em}
\noindent \textbf{Dimensionality Reduction.} 
It has been shown in many black-box attack methods \cite{DongSWLL0019, Huang020, IlyasEM19,GuoFW19} that searching or optimizing the adversarial perturbation in a suitable low-dimensional subspace can significantly improve query efficiency.
To further improve the query efficiency, here we also combine the dimensionality reduction technique with $\mathcal{CG}$-ATTACK. 
Specifically, We adopt the technique based on discrete cosine transform (DCT).
The general procedure of $\mathcal{CG}$-ATTACK with DCT is summarized in Algorithm \ref{alg:evolution}, where we adopt a popular variant of ES-based method, \ie, the co-variance matrix adaptation evolution strategy (CMA-ES) \cite{Hansen16a} as the basic algorithm. The details of DCT and the standard CMA-ES algorithm will be presented in Sec. 3 and 4 of the \textbf{Supplementary Material}, respectively.

\vspace{-.5em}
\section{Experiments}

\vspace{-.3em}
\subsection{Experimental Settings}
\label{sec: setting}
\vspace{-.1em}

\noindent \textbf{Datasets and Evaluation Metrics.}
Following the setting in \cite{DuZZYF20}, we choose 1,000 images randomly from the testing set of CIFAR-10 \cite{krizhevsky2009} and the validation set of 10 randomly selected classes from ImageNet \cite{RussakovskyDSKS15} for evaluation, respectively. 
For both datasets, we normalize the input to $[0, 1]$. The maximum distortion of adversarial images for CIFAR-10 is set as $\epsilon = 0.03125$ and for ImageNet is set as $\epsilon = 0.05$. The maximum number of queries is set to 10,000 for all the experiments.
As in prior works \cite{GuoYZ19, MoonAS19}, we adopt the attack success rate (ASR), the mean and median number of queries of successful attacks to evaluate the attack performance.

\vspace{0.2em}
\noindent \textbf{Target and Surrogate Models.} \label{surrogate} 
For CIFAR-10, we consider four target models: VGG-15 \cite{SimonyanZ14a}, ResNet-Preact-110 \cite{HeZRS16}, DenseNet-BC-110 \cite{HuangLMW17} and PyramidNet-110 \cite{HanKK17}.
The models are implemented based on the GitHub repository \footnote{\small \textit{https://github.com/hysts/pytorch\_image\_classification}}. 
Unless otherwise specified, we conduct the standard training on the training set of each dataset. 
The top-1 error rates of these four target models on the standard testing set of CIFAR-10 are (7.24\%, 10.04\%, 4.68\%, 7.24\%), respectively.
For ImageNet, we also evaluate our method on four target models: VGG-16 \cite{SimonyanZ14a}, ResNet-18 \cite{HeZRS16}, SqueezeNet \cite{IandolaMAHDK16} and GoogleNet \cite{SzegedyLJSRAEVR15}. The models are based on the official implementation of Pytorch and the pre-trained parameters are downloaded from torchvision.
The top-1 error rates of these target models on the validation set of ImageNet are (28.41\%, 30.24\%, 41.90\%, 30.22\%), respectively.
To further mitigate the possible negative effect due to the surrogate bias of model architectures, on each dataset, when attacking one target model, we treat the other three as surrogates. 

Besides, we also consider the attack against adversarially defended models. Following \cite{Huang020}, the defended models for CIFAR-10 were trained based on PGD adversarial training \cite{MadryMSTV18} and the SOTA models from \cite{XieWMYH19} are directly adopted for ImageNet. More specifically, ResNet50 and WResNet \cite{ZagoruykoK16} are adopted as the surrogate and target models for CIFAR-10 and ResNet152 Denoise and RexneXt101 Denoise from \cite{XieWMYH19} are adopted as surrogate and target models for ImageNet.

\begin{figure}[ht] 
\vspace{-.7em}
\begin{center}
\centerline{\includegraphics[width=0.75\linewidth]{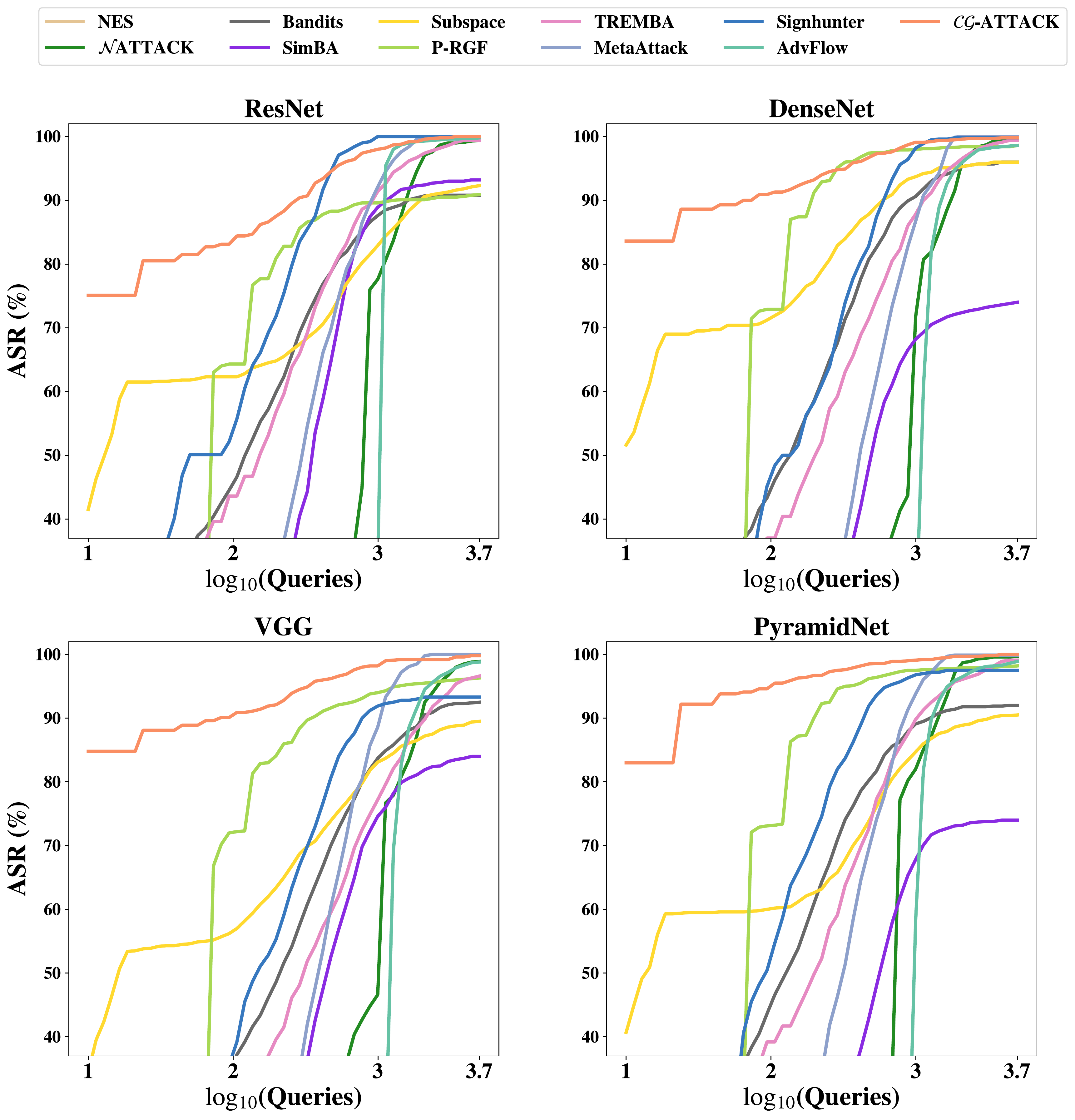}}
\caption{Attack success rate (ASR $\%$) $w.r.t.$ query numbers for untargeted attacks on CIFAR-10.}
\label{fig: query curve on CIFAR-10}
\vspace{-3em}
\end{center}
\end{figure}

\begin{table*}[t] 
\vspace{-0.3em}
\centering
\footnotesize
\caption{Attack success rate (ASR $\%$), mean and median number of queries of untargeted attack and targeted attack (target class $0$) on CIFAR-10. {The first 5 methods (from 'NES' to 'Signhunter') are pure query-based attacks, while the other methods are query-and-transfer-based attacks.} The best and second-best values among methods that achieve more than 90\% ASR are highlighted in bold and underline.}
\scalebox{0.72}{
\begin{tabular}{c p{.019\textwidth}<{\centering}p{.025\textwidth}<{\centering}p{.04\textwidth}<{\centering}p{.02\textwidth}<{\centering}p{.025\textwidth}<{\centering}p{.04\textwidth}<{\centering}p{.02\textwidth}<{\centering}p{.025\textwidth}<{\centering}p{.04\textwidth}<{\centering}p{.02\textwidth}<{\centering}p{.025\textwidth}<{\centering}p{.04\textwidth}<{\centering}
p{.002\textwidth}<{\centering}
p{.019\textwidth}<{\centering}p{.025\textwidth}<{\centering}p{.04\textwidth}<{\centering}p{.02\textwidth}<{\centering}p{.025\textwidth}<{\centering}p{.04\textwidth}<{\centering}p{.02\textwidth}<{\centering}p{.025\textwidth}<{\centering}p{.04\textwidth}<{\centering}p{.02\textwidth}<{\centering}p{.025\textwidth}<{\centering}p{.04\textwidth}<{\centering}}
\hline
 & \multicolumn{12}{c}{Untargeted Attack} & & \multicolumn{12}{c}{Targeted Attack}\\
  \cline{2-13}\cline{15-26}
 Target Model $\rightarrow$  & \multicolumn{3}{c}{ResNet} & \multicolumn{3}{c}{DenseNet} & \multicolumn{3}{c}{VGG} & \multicolumn{3}{c}{PyramidNet} & 
 & 
 \multicolumn{3}{c}{ResNet} & \multicolumn{3}{c}{DenseNet} & \multicolumn{3}{c}{VGG} & \multicolumn{3}{c}{PyramidNet}\\
 Attack Method $\downarrow$  & ASR    & Mean   & Median    & ASR     & Mean   & Median    & ASR   & Mean   & Median   & ASR     & Mean    & Median & & ASR    & Mean   & Median    & ASR     & Mean   & Median    & ASR   & Mean   & Median   & ASR     & Mean    & Median 
 \\
\hline
NES \cite{IlyasEAL18}     & 91.2 & 169.2 & 62.0      & 94.3 & 249.4 & 112.0        & 91.7 & 284.3 & 98.0       & 95.9 & 385.4 & 168.0  &  & 68.7 & 2973.5 & 1102.0      & 84.9 & 6932.4 & 4125.0      & 77.3 & 4192.4 & 2961.0       & 71.2 & 3977.8 & 2623.0      \\
 $\mathcal{N}$ATTACK \cite{LiLWZG19} & \underline{99.6} & 767.2 & 628.0      & 99.6 & 824.4 & 672.0        & 99.7 & 902.4 & 736.0       & \textbf{100.0} & 675.8 & 548.0  & & 99.1 & 1817.3 & 1548.0      & 100.0 & 1718.5 & 1493.0        & 100.0 & 3232.8 & 2874.0       & 100.0 & 1569.3 & 1288.0       \\
Bandits \cite{IlyasEM19}     & 90.8 & 193.4 & 88.0      & 96.0 & 206.3 & 96.0        & 93.0 & 361.5 & 158.0       & 92.0 & 194.9 & 92.0   &   & 72.6 & 3660.1 & 2812.0      & 80.0 & 4154.8 & 3842.0      & 83.4 & 3967.6 & 3860.0       & 77.8 & 4484.6 & 3876.0       \\
 SimBA \cite{GuoGYWW19}        & 93.2 & 432.1 & 235.0     & 74.0  &  480.5  &    223.0 & 68.3  &  632.3  &    237.0 & 84.0  &  455.5  &    270.0  &  & \textbf{100.0} & 940.0 & 885.0         & \textbf{100.0}  &  838.8  &    777.0 & \underline{99.5}  &  1343.2  &    1210.0 & \textbf{100.0}  &  \underline{865.8}  &    779.0 \\
 Signhunter \cite{Al-DujailiO20} & \textbf{100.0}  &  135.1  &    47.0 & \underline{99.8}  &  213.8  &    119.0 & 93.3  &    244.3 & 102.0   & 97.5  &    161.9 & 69.0 & & \textbf{100.0}  &  \underline{894.1}  &    \underline{657.0}   & \textbf{100.0}   &  \underline{826.9}  &    \underline{679.0}  & \textbf{99.7}  &  1431.7  &    1121.0 & \textbf{100.0}  &  1111.6  &    878.0     \\   
 Subspace \cite{GuoYZ19}   & 93.0 & 301.8 & \underline{12.0}      & 96.0  &  115.8 &    \underline{12.0}   & 90.0  &  272.0  &    \underline{12.0}  & 91.0  &  255.4  &    \underline{10.0} & & 78.0 & 2409.3 & 1630.0      & 94.0  &  1528.4 &    1012.0 & 67.0  &  2129.1  &    1366.0 & 80.0  &  2241.3  &    1586.0  \\
 P-RGF \cite{ChengDPSZ19}      & 92.2 & 121.8 & 62.0      & 99.6 & \underline{111.7} & 62.0        & 96.8 & 176.4 & 62.0   & \underline{98.2} & 135.8 & 62.0  &  & 70.6  & 1020.8  &    390.0  & 77.1 & 1037.1 & 438.0       & 61.3 & 1083.9  &   360.0    & 50.3  &  1108.8  &    436.0        \\
TREMBA \cite{Huang020}     & 90.9 & \underline{120.7} & 64.0      & 97.8 & 126.4 & 66.0        & 97.7 & \underline{125.5} & 63.0        & 97.9 & \underline{82.3} & 39.0 &   & 91.2  &  1125.3  &    868.0 & 92.3  &  1123.4  &    879.0 & 96.5  &  \underline{1331.5}  &    1142.0 & 98.1  &  1082.4  &    \underline{759.0}        \\
 MetaAttack \cite{DuZZYF20} & \textbf{100.0} & 363.2 & 153.0      & \textbf{100.0} & 411.5 & 225.0      & \textbf{100.0} & 392.0 & 161.0      & \textbf{100.0} & 320.4 & 191.0 &  & 98.7 & 1953.3 & 1537.0      & \underline{99.8} & 2013.7 & 1793.0      & 86.1 & 3045.6 & 2307.0       & \underline{98.9} & 2054.6 & 1665.0       \\
AdvFlow \cite{advflow} & 97.2 & 841.4 & 598.0      & \textbf{100.0} & 1025.3 & 736.0        & 98.2 & 1079.1 & 862.0       & 99.7 & 857.5  &  562.0 & & 98.6 & 911.7 & 822.0      &  96.3 & 1021.5 & 868.0   & 97.4 & 1144.1 & \underline{946.0}      & \textbf{100.0} & 908.1  &  824.0  \\
 $\mathcal{CG}$-ATTACK & \textbf{100.0}  &  \textbf{81.6}  &  \textbf{1.0}       & \textbf{100.0}  &  \textbf{43.3}  &  \textbf{1.0}         & \underline{99.9}  &  \textbf{56.4}  &  \textbf{1.0}        & \textbf{100.0}  &  \textbf{30.1}  &  \textbf{1.0} & & \underline{99.9}  &  \textbf{696.4}  &    \textbf{421.0}  & \textbf{100.0}  & \textbf{787.1}  &    \textbf{621.0}    & 98.8  &  \textbf{861.1}  &    \textbf{581.0}   & \underline{98.9}  &  \textbf{651.2}  &    \textbf{461.0}  \\        

\bottomrule
\end{tabular}
\label{cifar_unt}
}
\vspace{-1.5em}
\end{table*}

\vspace{0.2em}
\noindent \textbf{Compared methods.} 
Several SOTA score-based black-box attack methods are compared, including NES \cite{IlyasEAL18}, Bandits \cite{IlyasEM19},  $\mathcal{N}$ATTACK \cite{LiLWZG19}, SimBA \cite{GuoGYWW19}, Subspace \cite{GuoYZ19}, P-RGF \cite{ChengDPSZ19}, TREMBA \cite{Huang020}, MetaAttack \cite{DuZZYF20}, Signhunter \cite{Al-DujailiO20} and AdvFlow \cite{advflow}.
All of them are implemented using the source codes provided by their authors. 
\vspace{-.2em}
\subsection{Experiments on Closed-set Attack Scenario}\label{sec:experiments:closedset}
\vspace{-.2em}
\subsubsection{Performance of Black-box Attack on CIFAR-10}

\noindent \textbf{Untargeted Attack.} 
In this case, one attack is successful if the predicted class of the adversarial example is different from the ground-truth label.
The results are reported in the left half of Tab.~\ref{cifar_unt}.
It shows that the proposed $\mathcal{CG}$-ATTACK achieves 100\% ASR on ResNet, DenseNet, and PyramidNet, and 99.9\% ASR on VGG, which demonstrates the effectiveness of our method.
$\mathcal{CG}$-ATTACK is also very query-efficient. The mean number of queries is the lowest under all four target models in Tab. 1.  
More surprisingly, the median number of queries of $\mathcal{CG}$-ATTACK is just 1, which means that we successfully fool the target model with just one query for more than $50\%$ attacked images. It reveals that the c-Glow model pretrained on surrogate models is a good approximation to the perturbation distribution of the target model.
In contrast, the second-best median queries are obtained by Subspace \cite{GuoYZ19}, which are more than 10x of ours, and with much lower  ASR. 
The curves of the average ASR on all evaluation images $w.r.t$ the query number are shown in Fig.~\ref{fig: query curve on CIFAR-10}. It clearly highlights the superiority of our $\mathcal{CG}$-ATTACK method. Especially in the stage of low query numbers, $\mathcal{CG}$-ATTACK achieves very high ASR efficiently.

\noindent \textbf{Targeted Attack.} 
Following \cite{Huang020}, we conduct targeted attacks with three target classes, including 0 (airplane), 4 (deer), and 9 (truck). 
When attacking for one target class, images with the same ground-truth class are skipped. 
Due to space limitations, we report the attack results of the target class 0 in the right half of Tab.~\ref{cifar_unt}, and leave the results of the other two target classes in Sec. 6.1 of the \textbf{Supplementary Material}.
As shown in Tab.~\ref{cifar_unt}, our $\mathcal{CG}$-ATTACK method achieves at least 98.8\% ASR on all target models.
Besides, the mean and median query numbers of $\mathcal{CG}$-ATTACK are significantly lower than that of all compared methods, demonstrating its query efficiency.
Signhunter \cite{Al-DujailiO20} obtains a slightly higher ASR than $\mathcal{CG}$-ATTACK on VGG and PyramidNet, but with the cost of more than 1.6x query numbers.

\vspace{-0.9em}
\subsubsection{Performance of Black-box Attack on ImageNet}\label{sec:imagenet}
\vspace{-.5em}
We perform both targeted and untargeted attacks against models on the ImageNet dataset. We report the results for untargeted attacks and leave the results for targeted attacks in Sec. 6.2 of  the \textbf{Supplementary Material}.
The results are summarized in Tab.~\ref{tiny_unt}.
It shows that $\mathcal{CG}$-ATTACK performs better than compared methods in most cases.
Specifically, 
when attacking the GoogleNet model, $\mathcal{CG}$-ATTACK achieves the highest ASR with the lowest mean and median number of queries among all methods.
When attacking SqueezeNet, the Signhunter is slightly higher (0.7\% higher) than ours in terms of ASR, but its mean and median number of queries are 2.8x and 64x of ours. On ResNet and SqueezeNet, $\mathcal{CG}$-ATTACK achieves the best values of both mean and median number of queries.
Moreover, we also study the effect dimensionality reduction in Sec. 6.4 of the \textbf{Supplementary Material}.

\begin{table}[t] 
\centering
\small
\caption{Attack success rate (ASR $\%$), mean and median number of queries of untargeted attack on ImageNet. The best and second-best values among methods that achieve more than 90\% ASR are highlighted in bold and underline, respectively.}
\scalebox{0.6}{
\begin{tabular}{ p{.13\textwidth}<{\centering} p{.02\textwidth}<{\centering}p{.03\textwidth}<{\centering}p{.04\textwidth}<{\centering}p{.02\textwidth}<{\centering}p{.03\textwidth}<{\centering}p{.04\textwidth}<{\centering}p{.02\textwidth}<{\centering}p{.03\textwidth}<{\centering}p{.04\textwidth}<{\centering}p{.02\textwidth}<{\centering}p{.03\textwidth}<{\centering}p{.04\textwidth}<{\centering}}
\hline
 Target model $\rightarrow$ & \multicolumn{3}{c}{ResNet} & \multicolumn{3}{c}{GoogleNet} & \multicolumn{3}{c}{VGG} & \multicolumn{3}{c}{SqueezeNet} \\
 \cmidrule(lr){2-4}\cmidrule(lr){5-7}\cmidrule(lr){8-10}\cmidrule(lr){11-13}
 Attack Method $\downarrow$  & ASR    & Mean   & Median    & ASR     & Mean   & Median    & ASR   & Mean   & Median   & ASR     & Mean    & Median 
 \\
\hline
NES \cite{IlyasEAL18}     & 91.2 & 1642.1 & 664.0      & 86.3 & 1725.3 & 612.0       & 81.6 & 1394.7 & 586.0        & 87.5 & 1473.3 & 596.0      \\
 $\mathcal{N}$ATTACK \cite{LiLWZG19} & 95.3  & 1124.6 &   760.0 & 95.6  & 1266.4 &   864.0   & 90.9  & 874.6 &   692.0 & 94.8  & 1362.2 &   812.0   \\
Bandits \cite{IlyasEM19}     & 90.3 & 972.3 & 248.0      & 89.7 & 1247.1 & 462.0       & 84.3 & 991.3 & 773.0        & 88.2 & 1173.4 & 862.0      \\
SimBA \cite{GuoGYWW19}        & 96.7  & 577.3 &   245.0   & \underline{99.1}  & 995.0 &   382.0 & 93.4  & 882.6 &   382.0   & 94.3  & 1052.3 &   766.0 \\
Signhunter \cite{Al-DujailiO20} & \textbf{100.0}  & \underline{278.2} &   \underline{48.0}   & \textbf{100.0}  & 284.7 &   124.0    & \textbf{100.0}   &   218.9 & \underline{64.0}     & \textbf{100.0}  &   315.9 & \underline{72.0}      \\
Subspace \cite{GuoYZ19}   & 93.1  & 533.8 &   224.0    & 96.3  & 632.1 &  322.0    & 94.3  & 533.2 &   310.0    & 95.7  & 589.2 &   272.0   \\
P-RGF \cite{ChengDPSZ19}      & 96.1 & 528.1 & 284.0       & 97.3 & 466.2 & 271.0         & 97.3 & 336.1 & 184.0          & 94.7 & 463.7 & 172.0        \\
TREMBA \cite{Huang020}     & \textbf{100.0}  & 332.4 &   121.0  & 96.7  & \underline{246.6} &   \underline{101.0}    & 97.6  & \underline{196.2} &   81.0    & 97.3  & \underline{272.1} &   131.0   \\
MetaAttack \cite{DuZZYF20} & 94.8  & 335.2 &   167.0 & 96.3  & 288.6 &   121.0   & 93.6  & 311.2 &   96.0 & 96.3  & 288.3 &   132.0 \\
AdvFlow \cite{advflow}     & 96.7  & 746.1 &   482.0  & 99.3  & 694.8 &   364.0    & 95.5  & 1022.6 &   748.0    & 99.2  & 894.3 &  521.0   \\
$\mathcal{CG}$-ATTACK     & \underline{97.3}  & \textbf{210.4} &   \textbf{21.0}   & \textbf{100.0}  & \textbf{138.8} &   \textbf{21.0}    & \underline{99.4}  & \textbf{77.3} &   \textbf{1.0}    & \underline{99.3}  & \textbf{132.9} &   \textbf{21.0}  \\

\bottomrule
\end{tabular}
\label{tiny_unt}
}
\vspace{-1.4em}
\end{table}

\vspace{-1.0em}
\subsubsection{Black-box Attack on Defended Models}
\vspace{-0.7em}
In this section, we perform an untargeted attack against defended models based on adversarial training, and the results are reported in Tab. \ref{tab:defense}. In addition to the results based on defended surrogate models mentioned in Sec. \ref{surrogate} (listed as $\mathcal{CG}$-ATTACK-Robust), we also present the results for undefended surrogate models (listed as $\mathcal{CG}$-ATTACK). Note that for other baseline methods, only results for defended surrogate models are presented. From Tab. \ref{tab:defense}, we can see that even without the defended surrogate models, $\mathcal{CG}$-ATTACK still outperforms the baseline methods in all three metrics. This shows that our method is capable of efficiently adapting the conditional adversarial distribution despite the large surrogate biases in model architectures. Note that with a better surrogate model, \ie, the defended model, $\mathcal{CG}$-ATTACK-Robust consistently improves attack performance over $\mathcal{CG}$-ATTACK in terms of ASR (5\% higher), mean (7\% lower), and median queries (5\% lower).

\begin{table}[t]
    \centering
    \setlength\tabcolsep{4.0pt}
    \caption{Attack success rate (ASR $\%$), mean and median number of queries of black-box untargeted attack on defended CIFAR10 and ImageNet model. The best and second-best values among methods are highlighted in bold and underline, respectively.}
    \scalebox{0.75}{
    \begin{tabular}{ccccccc}
        \hline
        {Target model$\rightarrow$} & 
        \multicolumn{3}{c}{CIFAR10 WResnet} & \multicolumn{3}{c}{ImageNet RexneXt101} \\
        \cmidrule(r){2-4} \cmidrule(r){5-7}
        {Attack method$\downarrow$} & ASR & Mean & Median & ASR & Mean & Mean \\ \hline
        NES \cite{IlyasEAL18}   & 13.2 & 5682.1 & 2261.3 &  10.3 & 7745.2 & 3943.0 \\
         $\mathcal{N}$ATTACK \cite{LiLWZG19}  & 26.1  & 4753.9 & 2763.0 &  29.7   & 6352.4   & 3971.0   \\
        Bandits \cite{IlyasEM19}   & 18.7 & 3127.5 & 1263.2 &  16.4 & 4962.3 & 3138.0 \\
        SimBA \cite{GuoGYWW19}  & 29.6 & 3826.9 & 2642.0  &  25.7 & 7152.6 & 3072.0 \\
        Signhunter \cite{Al-DujailiO20} & 58.1  & 986.1 &  583.0 & 60.1   & 1585.3    &   769.0 \\
        Subspace \cite{GuoYZ19}   & 31.3 & 3965.7 & 2492.0 &  26.1 & 6973.2 & 4175.0 \\
        P-RGF \cite{ChengDPSZ19}      & 22.9 & 4983.2 & 3617.0 & 21.2 & 7791.4 & 5823.0  \\
        TREMBA \cite{Huang020}     & 56.2 & 1242.4 & 726.0 &  51.3 & 3952.0 & 1944.0 \\
        MetaAttack \cite{DuZZYF20} & 47.1  & 1527.6 & 681.0 &  46.5 & 2823.7 & 1149.0 \\
        AdvFlow \cite{advflow} & 36.8  & 2386.2 & 1124.0  &  32.7   & 4952.8 & 3168.0  \\
        \hline
        {$\mathcal{CG}$-ATTACK} & \underline{58.5} & \underline{789.7} & \underline{371.0} & \underline{63.3} & \underline{1374.0} &  \underline{621.0} \\
        {$\mathcal{CG}$-ATTACK-Robust} & \textbf{64.3} & \textbf{606.1} & \textbf{341.0} & \textbf{72.1} & \textbf{1305.1} & \textbf{581.0} \\
        \hline
    \end{tabular}
    }
    \label{tab:defense}
    \vspace{-1.0em}
\end{table}

\begin{table*}[tbh] 
\centering
\footnotesize
\caption{Attack success rate (ASR $\%$), mean and median number of queries of open-set untargeted attack on CIFAR-10 (\textbf{Case 1} and \textbf{Case 2}). {The first 5 methods (from 'NES' to 'Signhunter') are pure query-based attacks, while the other methods are query-and-transfer-based attacks.} The best values among methods are highlighted in bold.}
\scalebox{0.72}{
\begin{tabular}{c p{.019\textwidth}<{\centering}p{.025\textwidth}<{\centering}p{.04\textwidth}<{\centering}p{.02\textwidth}<{\centering}p{.025\textwidth}<{\centering}p{.04\textwidth}<{\centering}p{.02\textwidth}<{\centering}p{.025\textwidth}<{\centering}p{.04\textwidth}<{\centering}p{.02\textwidth}<{\centering}p{.025\textwidth}<{\centering}p{.04\textwidth}<{\centering}
p{.002\textwidth}<{\centering}
p{.019\textwidth}<{\centering}p{.025\textwidth}<{\centering}p{.04\textwidth}<{\centering}p{.02\textwidth}<{\centering}p{.025\textwidth}<{\centering}p{.04\textwidth}<{\centering}p{.02\textwidth}<{\centering}p{.025\textwidth}<{\centering}p{.04\textwidth}<{\centering}p{.02\textwidth}<{\centering}p{.025\textwidth}<{\centering}p{.04\textwidth}<{\centering}}
\hline
 & \multicolumn{12}{c}{Case 1} & & \multicolumn{12}{c}{Case 2}\\
 \cline{2-13}\cline{15-26}
 Target Model $\rightarrow$  & \multicolumn{3}{c}{ResNet} & \multicolumn{3}{c}{DenseNet} & \multicolumn{3}{c}{VGG} & \multicolumn{3}{c}{PyramidNet} &
 &
 \multicolumn{3}{c}{ResNet} & \multicolumn{3}{c}{DenseNet} & \multicolumn{3}{c}{VGG} & \multicolumn{3}{c}{PyramidNet}\\
 Attack Method $\downarrow$ & ASR    & Mean   & Median    & ASR     & Mean   & Median    & ASR   & Mean   & Median   & ASR     & Mean    & Median & & ASR    & Mean   & Median    & ASR     & Mean   & Median    & ASR   & Mean   & Median   & ASR     & Mean    & Median 
 \\
\hline
 NES \cite{IlyasEAL18}     & 93.1 & 225.2 & 79.0      & 96.7 & 188.5 & \underline{68.0}       & 93.2 & \underline{144.6} & 81.0        & 97.2 & 273.5 & 133.0  & & 94.4 & 251.3 & 101.0      & 92.9 & 373.4 & 186.0       & 94.2 & 343.0 & 192.0        & 96.3 & 309.1 & 214.0    
 \\
 $\mathcal{N}$ATTACK \cite{LiLWZG19} & \underline{99.7}  & 688.3 &  264.0 & 99.3  & 645.2 &   262.0   & 99.1  & 725.7 &  318.0 & 98.6  & 643.0 &  205.0 & & 98.2  & 607.2 &  335.0 & \textbf{99.1}  & 706.7 &   349.0   & \underline{98.8}  & 876.0 &  705.0 & \textbf{99.3}  & 731.9 &  581.0   
 \\
 Bandits \cite{IlyasEM19}     & 91.7 & 167.1 & 67.0      & 94.5 & 178.2 & 83.0       & 94.6 & 287.5 & 112.0        & 96.9 & 212.1 & 78.0 &   & 92.5 & 226.3 & 98.0      & 93.6 & 154.1 & 62.0       & 95.5 & 249.3 & 174.0        & 93.2 & 163.3 & 64.0      \\
 SimBA \cite{GuoGYWW19}        & 91.2  & 386.7 &   168.0   & {84.5}  & 297.6 &  173.0 & 73.2  & 319.4 &  163.0   & 85.1  & 413.9 &   258.0  &   & 93.6  & 332.0 &   101.0   & {91.2}  & 287.3 &  121.0 & 88.4  & 427.9 &  226.0   & 90.2  & 378.3 &   215.0 \\
 Signhunter\cite{Al-DujailiO20} & \textbf{100}  &  \underline{141.3}  &   \underline{ 41.0} & \underline{99.8}  &  156.3  &    \underline{68.0} & \textbf{100} &    168.1 & 84.0   & \underline{98.9}  & 182.3 & \underline{77.0}  & &  \textbf{100}  &  \underline{137.8}  &    \underline{41.0} & \textbf{99.1}  &  149.8  &    47.0 & \textbf{100} &    166.4 & 69.0   & 98.7  & \underline{143.5} & \underline{53.0}    \\    
Subspace \cite{GuoYZ19}   & 92.2  & 263.0 &   116.0    & 92.7  & 164.3 &  71.0    & 96.6  & 239.7 &  102.0    & 94.3  & 285.2 &   162.0 &  & 91.4  & 317.2 &   231.0    & 90.8  & 229.8 &  105.0    & 93.9  & 319.8 &  174.0    & 92.7  & 241.9 &   102.0    \\
 P-RGF \cite{ChengDPSZ19}   & 87.5 & 144.3 & 97.0      & 93.7 & {165.8} & 98.0        & 94.2 & 188.6 & 81.0        & {94.1} & 177.4 & 101.0 &    & 91.1 & 146.0 & 75.0      & 93.6 & 196.4 & 37.0        & 92.6 & \underline{152.5} & \underline{48.0}        & 92.9 & 125.3 & 66.0       \\
 TREMBA \cite{Huang020}   & 90.4 & 247.2 & 142.0      & 93.1 & \underline{148.9} & 96.0        & 95.1 & 196.1 & \underline{73.0}        & 93.2 & \underline{143.5} & 81.0    &   & 91.3 & 189.4 & 91.0      & 93.2 & 174.5 & 71.0        & 94.5 & 226.2 & 161.0        & 92.1 & 168.3 & 101.0           \\
 MetaAttack \cite{DuZZYF20} & 95.2 & 414.3 & 161.0      & 96.5 & 379.7 & 241.0      & \underline{98.3} & 427.2 & 201.0      & 96.4 & 364.8 & 151.0 &  & 94.7 & 386.4 & 201.0      & 93.2 & 425.9 & 361.0      & 93.8 & 362.4 & 161.0      & 95.1 & 374.3 & 191.0   \\
 AdvFlow \cite{advflow} & 94.3 & 682.9 & 411.0      & 99.3 & 1269.2 & 841.0        & 95.3 & 1165.3 & 841.0       & 93.2 & 963.1  &  587.0 &  & 93.1 & 788.1 & 473.0      & 94.9 & 885.3 & 624.0        & 92.8 & 1299.2 & 806.0       & 95.7 & 1092.8  & 784.0   \\
 $\mathcal{CG}$-ATTACK & \textbf{100}  &  \textbf{123.4}  &  \textbf{21.0}       & \textbf{100}  &  \textbf{88.5}  &  \textbf{1.0}         & {98.2}  &  \textbf{127.9}  &  \textbf{41.0}        & \textbf{99.1}  &  \textbf{61.1}  &  \textbf{1.0} &  & \underline{98.8}  &  \textbf{103.5}  &  \textbf{21.0}       & \underline{98.2}  &  \textbf{132.3}  &  \textbf{21.0}         & {98.4}  &  \textbf{136.8}  &  \textbf{21.0}        & \underline{99.2}  &  \textbf{109.6}  &  \textbf{21.0}\\        
\bottomrule
\end{tabular}
\label{disjoint}
}
\vspace{-1.5em}
\end{table*}
\vspace{-0.3em}
\subsection{Experiments on Open-set Attack Scenario}\label{sec:experiments:openset}
\vspace{-0.3em}
\subsubsection{Black-box Attack on Benchmark Datasets}
\vspace{-0.3em}
In Sec. \ref{sec:experiments:closedset}, we have considered the \textbf{closed-set attack scenario} where surrogate and target models share the same training set, which has been widely adopted in many previous black-box attack methods \cite{GuoYZ19, ChengDPSZ19, Huang020, DuZZYF20}. 
However, in real-world scenarios, the attacker may not know the dataset set used for training the target model, dubbed \textbf{open-set attack scenario}.\comment{Consequently, surrogate models can only be trained using a different set with the target model. In this case, the effect of adversarial transferability between surrogate and target models should be further evaluated.} 
Specifically, we consider the following two cases. 
\textbf{Case 1}: the surrogate and target models are trained on disjoint images from same classes. In this case, the attacker has access to the class labels of the target training set, and creates a proxy training set by collecting images of each class from the internet. In our experiments, we evenly split the training images of each class, and surrogate models are trained on one half, while the target model on the other. 
\textbf{Case 2}: the surrogate and target models are trained on disjoint images from disjoint classes. In this case, the complete class labels are not released, and it is more challenging for the attacker to construct a similar proxy dataset to train surrogate models. Here we consider an extreme setting that the class labels of training sets used for training surrogate and tergate models are disjoint. Specially, we split the whole training set by classes evenly, and train surrogate models on one half, while target models on the other. 

We report the results on CIFAR-10 in Tab. \ref{disjoint} for both Case 1 and Case 2. 
Due to the space limit, the results on ImageNet will be presented in Sec. 6.3 of the \textbf{Supplementary Material}. 
As shown in the left half of Tab. \ref{disjoint}, $\mathcal{CG}$-ATTACK achieves the best values for ASR, mean and median number of queries when attacking ResNet, DenseNet and PyramidNet. When attacking VGG, Signhunter is sightly better in terms of ASR, but its mean and median query numbers are 1.3x and 2x of ours. For case 2, $\mathcal{CG}$-ATTACK achieves the lowest mean and median number of queries in all the categories with ASR of at least 98.2\%. Due to surrogate biases, the query-and-transfer based methods achieve lower ASR compared to the results in Tab. \ref{cifar_unt}. $\mathcal{CG}$-ATTACK, however, adapts well to the difference in training-set and obtain the least drop of only 0.53\% in ASR, which is the best result among these methods. Thus, the differences on training images and class labels will not significantly degrade the effect of adversarial transferability in $\mathcal{CG}$-ATTACK, making the proposed method more practical in real world scenarios.  
\vspace{-0.5em}
\subsubsection{Black-box Attack against Real-World API}
\vspace{-.3em}
To demonstrate the effectiveness of our attack against real-world systems, we further evaluate $\mathcal{CG}$-ATTACK by attacking the Imagga Tagging API\footnote{https://imagga.com/solutions/auto-tagging}, where the tagging model is trained with a unknown dataset of over 3000 types of daily life objects. This API will return an list of relevant labels associated with confidence scores for each query. 
We randomly selected 20 images from the ImageNet validation set for evaluation and set the query limit to 500. We define an untargeted attack aiming to remove original top-3 labels from the returned list, by minimizing the maximal score of these three labels. 
We pre-train the c-Glow model with four surrogate models on ImageNet, as described in Sec. \ref{sec:imagenet}. 
As shown in Tab. \ref{API}, our attack achieves a significantly higher ASR and lower number of queries than compared methods. 
Note that this attack belongs to the challenging Case 2 of open-set scenario, since the training sets for surrogate and target models are disjoint. 
It further verifies the effectiveness of $\mathcal{CG}$-ATTACK in real-world scenarios. 

\vspace{-0.3em}
\subsection{Summary of the Experimental Results}
In above closed-set experiments, our method achieves 35 best results and 6 second best results, out of the total 42 evaluation results. These results indicate 
In above open-set experiments, our method achieves 22 best results and 3 second best results out of the total 27 evaluations. Notably, we achieve significantly higher ASR and lower queries numbers than compared methods, when attacking real-world API. All above results support that our partial transfer mechanism can mitigate the possible negative effect of surrogate biases, thus boosting the black-box attack performance.

\comment{In all above results summarized in Tables~\ref{cifar_unt}-\ref{tiny_unt}, and there are 48 evaluation results in total. Among these results, our $\mathcal{CG}$-ATTACK method obtains 36 best and 6 second-best results. It fully demonstrates the superior performance of $\mathcal{CG}$-ATTACK on both effectiveness and efficiency, to all compared methods. 
Moreover, $\mathcal{CG}$-ATTACK always achieves the lowest median numbers of queries (except the targeted attack on VGG of Tiny-ImageNet), and even 1 at 4 results.  
It reflects that the search distribution $\bm{\pi}$ is very close to the intrinsic perturbation distribution of the target model, due to the powerful flexibility of the c-Glow model coupled with the Gaussian distribution, as well as the good transferability of the c-Glow model pretrained on surrogate models. 
}

\vspace{-0.4em}
\subsection{Supplementary Material}
\vspace{-0.5em}
Due to the space limit, some important experimental evaluations will be presented in the Supplementary Material, including: implementation details of $\mathcal{CG}$-ATTACK,  additional results of targeted attacks on both CIFAR-10 and ImageNet, additional results of open-set attacks on ImageNet, ablation studies about the effects of the c-Glow model and its initialization, verification of Assumption \ref{assumption 1}, as well as potential usage of the CAD approximated by c-Glow. These results could further highlight the superiority of $\mathcal{CG}$-ATTACK, and provide a deeper understanding of our method.

\begin{table}[t] 
\centering
\vspace{-0.5em}
\small
\caption{Attack success rate (ASR \%), mean and median number of queries of untargeted attack against \emph{Imagga} tagging API. The best results are highlighted in bold.}
\scalebox{0.75}{
\begin{tabular}{p{.035\textwidth}<{\centering}p{.05\textwidth}<{\centering}p{.08\textwidth}<{\centering}p{.08\textwidth}<{\centering}p{.07\textwidth}<{\centering}p{.07\textwidth}<{\centering}p{.07\textwidth}<{\centering}}
\hline
  & \scalebox{0.8}{NES\cite{IlyasEAL18}} & \scalebox{0.8}{$\mathcal{N}$ATTACK\cite{LiLWZG19}} & \scalebox{0.8}{Bandit\cite{IlyasEM19}} 
  & \scalebox{0.8}{SimBA\cite{GuoGYWW19}} &  \scalebox{0.8}{Signhunter\cite{Al-DujailiO20}} &
  \scalebox{0.8}{Subspace\cite{GuoYZ19}} \\
 \hline
 ASR & 30.0 & 50.0 & 50.0 & 70.0 &  50.0 & 55.0  \\
 Mean & 373.2 & 146.5 & 177.5 & 155.3 &  127.5 & 274.2 \\
 Median & 361.0 & 61.0 & 127.0 & 96.0 &  71.0 & 215.0  \\
\hline
& \scalebox{0.8}{P-RGF\cite{ChengDPSZ19}} &
  \scalebox{0.8}{TREMBA\cite{Huang020}} & \scalebox{0.8}{MetaAttack\cite{DuZZYF20}} &
  \scalebox{0.8}{AdvFlow\cite{advflow}} & \scalebox{0.8}{$\mathcal{CG}$-ATTACK} \\
  \hline
  ASR & 45.0 & 65.0 & 40.0 & 35.0 & \textbf{85} \\
  Mean & 194.3 & 87.2 & 312.4 & 368.6 & \textbf{75.7} \\
  Median & 128.0 & 51.0 & 182.0 & 143.0 & \textbf{21.0}\\
\bottomrule
\end{tabular}
\label{API}
}
\vspace{-1.2em}
\end{table}
\vspace{-.5em}
\section{Conclusion}
\vspace{-.5em}

This work presented a novel score-based black-box attack method, called $\mathcal{CG}$-ATTACK. The main idea is developing a novel mechanism of adversarial transferability that is robust to surrogate biases. More specifically, we proposed to transfer only partial parameters of CAD of surrogate models, while the remaining parameters are adjusted based on the queries to the target model on the attacked sample. We utilized the powerful c-Glow model to accurately model the CAD, and developed a novel efficient learning method based on randomly sampled perturbations.
\comment{To avoid additional queries on the target model, we proposed to firstly pretrain c-Glow on surrogate white-box models, using a novel efficient learning method based on randomly sampled perturbations, rather than adversarial perturbations. Then, we developed a novel transfer that only the mapping parameters of c-Glow are transferred from the pretrained c-Glow model, while the Gaussian parameters are automatically adjusted according to the attacked benign example and the query feedback returned by the target model. Consequently, the proposed attack method takes advantage of both adversarial transferability and queries.
}
Extensive experiments of attacking four DNN models on two benchmark datasets in both closed-set and open-set scenarios, as well as attack against real-world API, have fully verified the superior attack performance of the proposed method. 
%

\comment{
In this work, we proposed a novel search distribution in the evolution strategy (ES) method for solving the score-based black-box attack problem, based on the conditional Glow model coupled with the Gaussian distribution. This novel search distribution is flexible to capture the intrinsic distribution of adversarial perturbations conditioned on different benign examples. %
Besides, we proposed to pretrain the c-Glow model by approximating an energy-based model for the perturbation distribution of surrogate models. The pretrained c-Glow model is then used as initialization in ES for attacking the target model. 
Consequently, the proposed $\mathcal{CG}$-ATTACK method takes advantage of both query-based and transfer-based attack methods, to obtain high attack success rate and high efficiency simultaneously. 
Extensive experiments of attacking four models on two benchmark datasets have fully verified the superior attack performance of the proposed method, compared to several state-of-the-art methods. 
}

{\small
\bibliographystyle{ieee_fullname}
\bibliography{egbib}
}

\end{document}